# Contact resistance between carbon nanotubes

Alper Buldum and Jian Ping Lu
*Department of Physics and Astronomy, The University of North Carolina at Chapel Hill, Chapel Hill, NC, 27599*

Quantum transport properties of intermolecular nanotube contacts are investigated. We find that atomic structure in the contact region plays important roles and resistance of contacts varies strongly with geometry and nanotube chirality. Nanotube end-end contacts have low resistance and show negative differential resistance (NDR) behavior. Contact resistance can be dramatically decreased by exerting small pressure/force between the tubes if the contact is commensurate. Significant variation and nonlinearity of contact resistance may lead to new device applications.

Carbon nanotubes are candidates for future nanoelectronic device applications because of their outstanding electronic and transport properties [1–10]. Recently, single-electron transistors [4,5], field effect transistors [6] and rectifying diodes [3] have been reported. Intramolecular and intermolecular nanotube devices have been demonstrated [8–10]. To further enable such nanoelectronic devices and assembles it is clear that nanotube-nanotube contacts will play important roles. In this letter we report our studies on quantum transport properties of intermolecular nanotube contacts. We study the effect of tube-tube interaction on the transport properties of nanotubes for different positions, orientations and chiralities. Conductance and current-voltage (I-V) characteristics of both two and four-terminal nanotube contacts are investigated. We find that atomic structure in the contact region is crucial and resistance of contacts varies strongly with geometry and nanotube chirality. Nanotube end-end contacts in parallel has low resistance and shows Negative Differential Resistance (NDR) behavior. Four terminal cross-junctions have interesting transport properties which depend on the atomic structure in the contact region. Structural relaxation and applying external pressure/force decrease contact resistance dramatically if the contact is commensurate. Manipulation of these junctions enables significant variation of resistance and gives us different possibilities for device applications.

The electronic structure and the interaction between nanotubes are modeled by using $\pi$-orbital tight binding Hamiltonian [11,12]. Landauer-Büttiker formalism is used to calculate the conductance and the I-V characteristics with the surface Green's function matching method [13,14]. In this formalism the current on terminal $i$ can be written as [15]

$$I_i = \frac{2e}{h}\int_{-\infty}^{\infty} \bar{T}_{ij}(E,V)[f_i(E) - f_j(E)]dE \qquad (1)$$

where $\bar{T}_{ij}(E,V)$ is the transmission coefficient from terminal $i$ to $j$ and $f_i(E)$ is the Fermi function for terminal $i$. In the presence of applied bias, the energy levels are shifted and $\bar{T}_{ij}(E,V)$ is altered.

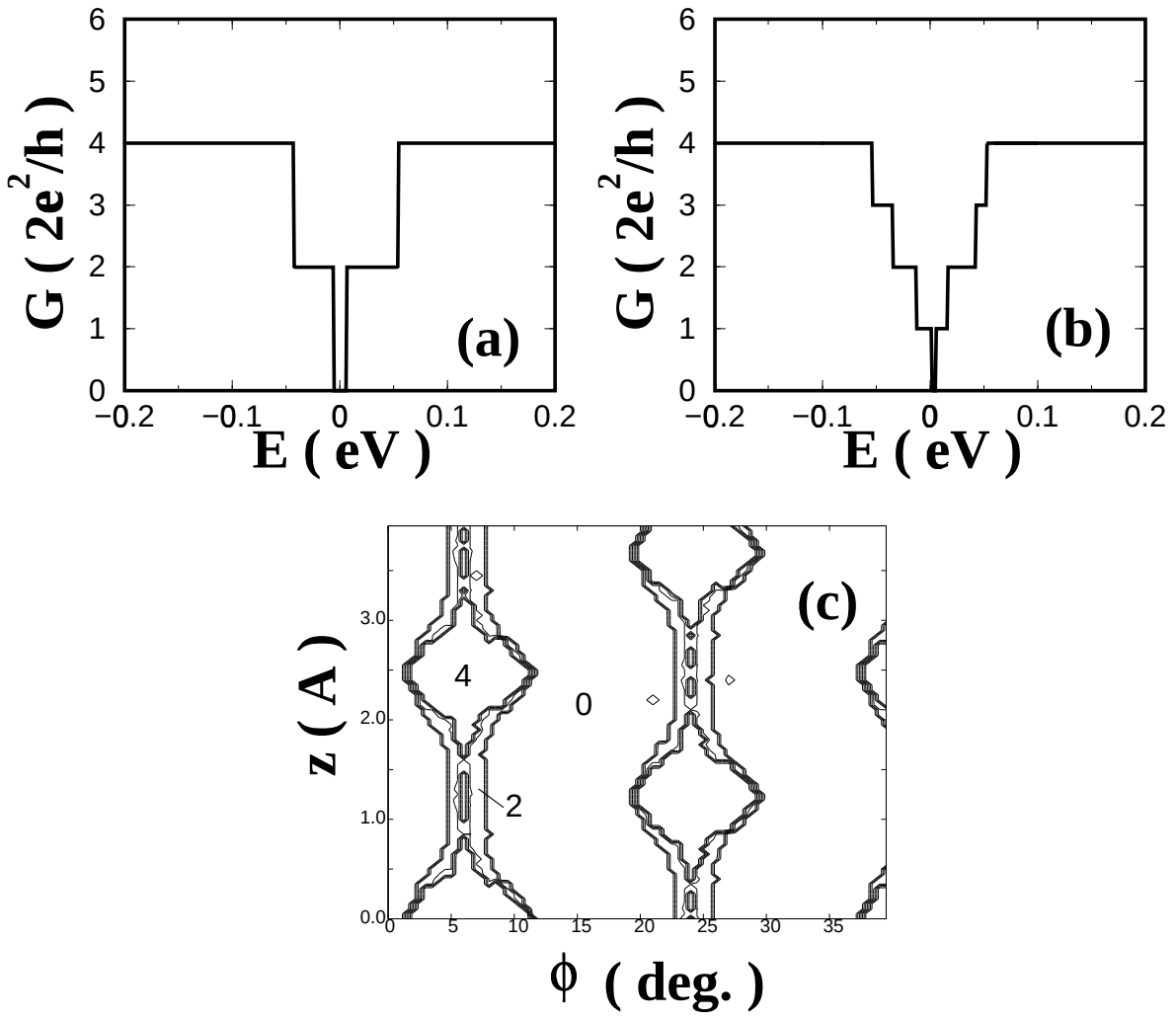

FIG. 1. (a) Conductance of two parallel armchair ((10,10)) tubes as a function of energy. The tubes are infinitely long with a 3.1 Å distance between each other. The leads are assumed to be connected to both of the tubes. (b) Conductance of two parallel zigzag ((18,0)) tubes. (c) Contour plot of conductance for two parallel (10,10) tubes as a function of translation $z$ along the tube's axis and spinning (rotation with respect to the tube's axis) ,$\phi$, of the second tube. Conductance has only values of 0,2 and 4.

When two nanotubes are brought together interaction between the nanotubes modify the electronic and transport properties. Thus, tube-tube interaction leads to variation of quantum conductance which depends on chirality and atomic structure in the contact region. Equilibrium positions are found for two parallel armchair and zigzag nanotubes and the contact region is commensurate like A-B stacking of graphite. Conductance of these two parallel armchair and zigzag nanotubes are presented in figure 1(a) and 1(b). Due to interaction between the tubes symmetry is broken and an energy gap is opened. An interesting result we find is that conductance of two parallel armchair tubes is quantized in units of $4e^2/h$ while that of zigzag tubes is quantized in units of $2e^2/h$. Thus, intertube interaction may be responsible for the single quantum conductance observed in the nanotube experiments [16]. Variation of conductance with atomic structure of contact region is further investigated by translating or spinning one of the tubes. In figure 1(c) the variation of conductance with translation, $z$, and spinning angle ,$\phi$, is presented. When contact region is commensurate like A-B stacking of Graphite conductance is zero since a gap is opened at the Fermi level.

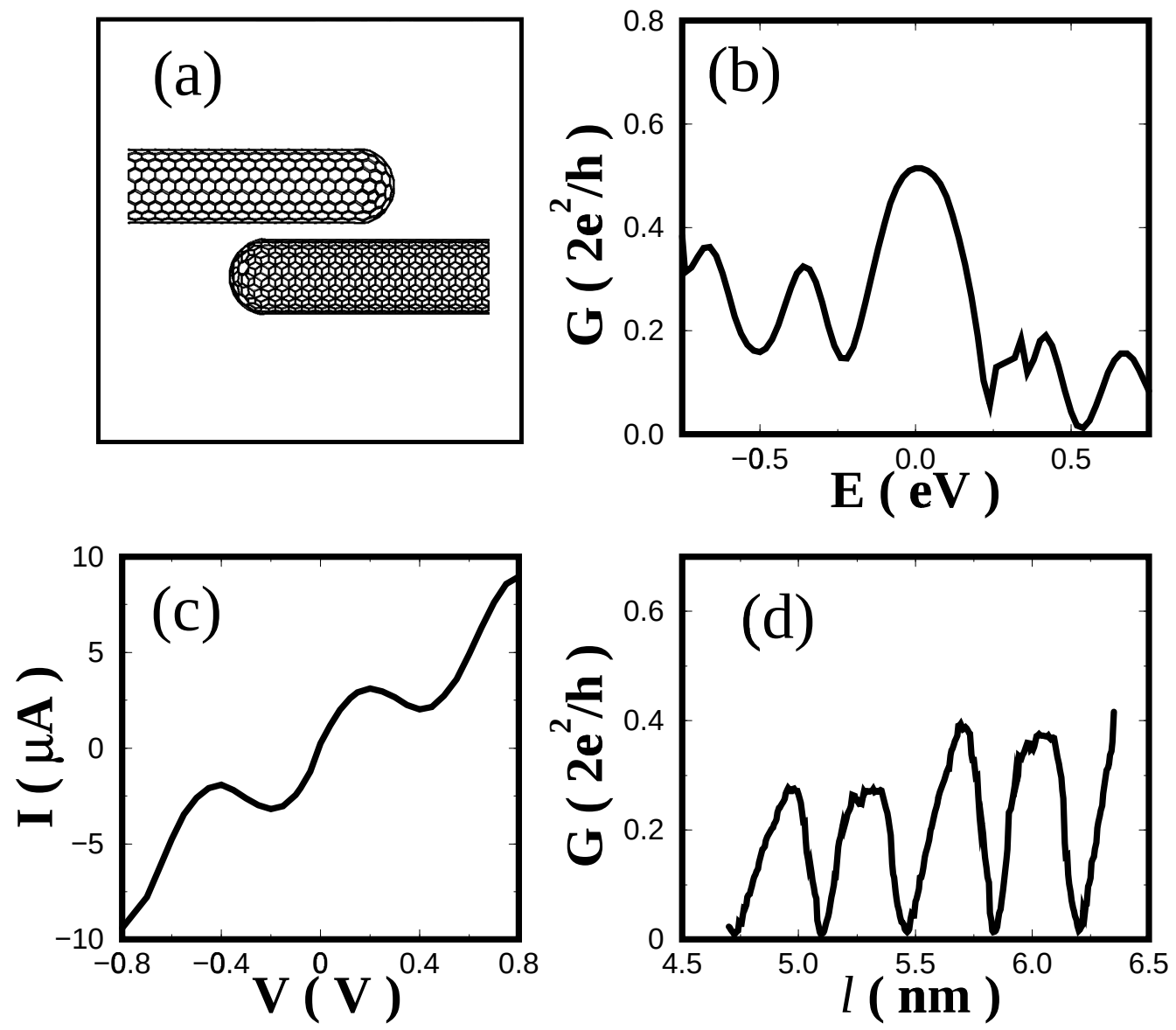

FIG. 2. (a) Two-terminal nanotube junction formed by connecting two (10,10) tubes in parallel and pointing in opposite directions. (b) The conductance, $G$, of two arm-chair tube ((10,10)-(10,10)) junction as a function of energy, $E$. The contact length is $l$ = 6.4 nm. Interference of electron waves yields resonance structure. (c) Current-voltage characteristics of a metal-metal ((10,10)-(10,10)) junction for $l$ = 4.6 nm. (d) The variation of conductance at the Fermi energy as a function of contact length, $l$ for (10,10)-(10,10) junction.

The simplest form of two-terminal nanotube contact is constructed by bringing two tube's end together (fig. 2(a)). Since the contact ( or interaction ) region is finite the junction shows quantum-interference effects. The interference of waves transmitted and reflected from the ends of the tubes yields resonance structure in conductance (fig 2(b)). The number of resonances increases by increasing the contact region length, $l$. This quantum-interference effect introduces Negative Differential Resistance (NDR) in the current-voltage characteristics which is shown in Fig. 2(c). NDR has many applications including high-speed switching, memory and amplification [17]. Another interesting feature is the nonlinear and quasi-periodic dependence of conductance on the contact length, $l$ (fig. 2(d)). Each pair of peaks form a period with length $3a_o$ ( $a_o$ = 2.46 Å unit cell length of armchair tubes ) which is the Fermi wavelength for armchair tubes. This periodicity is the same with the periodicity found in earlier experiments and theoretical calculations on finite nanotubes [18]. Notice that the conductance of this parallel end-end contact is high and comparable to the conductance of ideal tubes. Thus, this simple end-end contact geometry can be a preferred way of connecting multiple tubes in device applications. When the tubes are not parallel these features still exist but conductance is reduced by an order of magnitude.

An interesting four-terminal junction can be formed by placing one nanotube on top of another [10]. Multi-probe measurements can be performed on this junction with current passing two terminals and voltage measured using the other two. We find that conductance between the tubes is higher when two tubes are in-registry and the contact region is commensurate, where atoms from one tube are placed on top of another like A-B stacking of graphite. Thus, two perpendicular armchair tubes are out-of-registry in their cross-junction and atoms in the contact region are incommensurate. In this case, the resistance between the tubes are high. In contrast an armchair tube crossing a zigzag tube forms an in-registry junction and the resistance is low. In general, different transport properties can be achieved by manipulating these junctions such as rotating or translating one of the tubes with respect to the other. In figure 3(a) and 3(b) the variation of contact resistance with respect to rotation angle, $\Theta$, between the tubes is presented. The tubes are considered to be rigid and the upper tube (indiced 2-4) is rotated with an angle ,$\Theta$, which is between two tubes axes. We assumed that current is passing between terminals 1 and 4 and voltage is measured between 2 and 3. In the first junction ((18,0)-(10,10)) the tubes are in-registry at $\Theta$ = 30,90,150$^o$. In the second junction ((10,10)-(10,10)) the tubes are in-registry at $\Theta$ = 0,60,120,180$^o$. Large variation of resistance is observed by rotation and lower resistance values are found for in-registry orientations. Even when the tubes are in-registry the contact resistance can be different at different $\Theta$ due to change in the contact area. For example the resistance is lower at $\Theta = 30^o$ than at $\Theta = 90^o$ in (18,0)-(10,10) junction as the contact area at $\Theta = 30^o$ is larger. In figure 3(c), the variation of resistance with the translation of upper tube is shown for (18,0)-(10,10) junction. The upper tube ((18,0)) is translated along the lower tube's axis (x-direction) or perpendicular to the lower tube's axis (y-direction. Resistance values are represented with solid and dashed lines for translation along y- and x-directions, respectively. The variation of resistance is relatively small comparing with the variation with rotation but periodic with structural periodicity ( 2.46 Å for x-direction and 4.26 Å for y-direction). The lowest resistance is achieved when the contact structure is like A-A stacking of graphite. The significant variation of transport properties we found here is similar to the large dependence of mechanical/frictional properties on atomic scale registry [19,20].

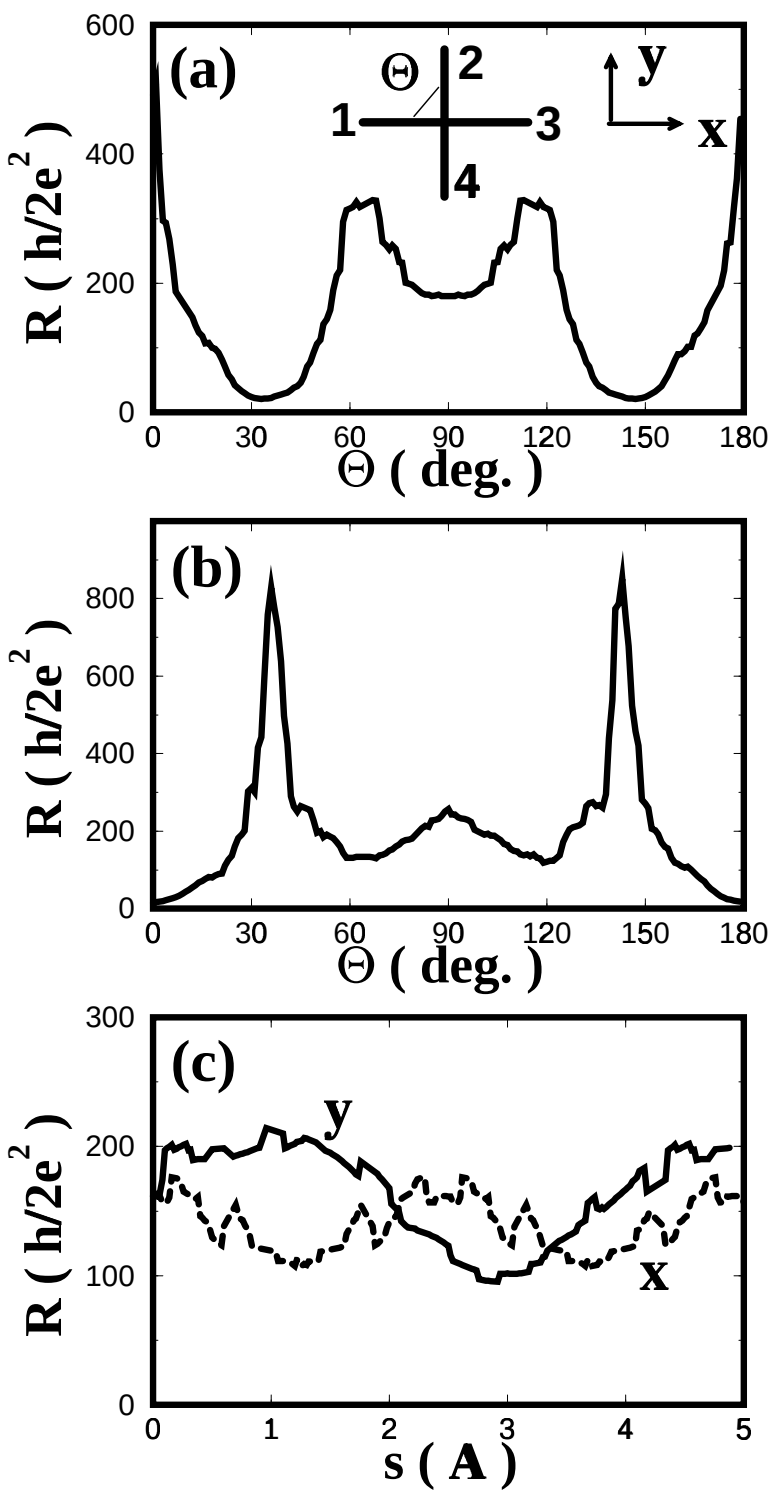

FIG. 3. Resistance of a four-terminal nanotube junction as a function of tube rotation and translation. The cross-junction model, rotation angle, Θ and the terminal indices are shown in the inset. The tube which is labeled by 2 and 4 is rotated by Θ or translated by $s$ in x- or y-directions. The current is passing between 1 and 4 and voltage is measured between 2 and 3. (a) Resistance of (18,0)-(10,10) junction as a function of Θ. The tubes are in-registry at $\Theta = 30, 90, 150^o$. (b) Resistance of (10,10)-(10,10) junction as a function of Θ. The tubes are in-registry at $\Theta = 0, 60, 120, 180^o$. (c) Resistance of (18,0)-(10,10) junction as functions of translation. The upper tube ((18,0), labeled by 2-4) translated along the lower tube's axis (x-direction) or perpendicular to the lower tube's axis (y-direction). Variation of resistance for translation in x- and y-directions are represented by dashed and solid lines, respectively.

In experimental systems, the junctions are on a substrate and electronic contact can be significantly enhanced by the structural relaxation of the tubes and adhesion between tubes and the substrate. We investigate the effect of relaxation by performing molecular dynamics simulations using empirical potentials [21]. The cross-junction is relaxed on a rigid surface [22] (fig. 4(a)), and constant forces ( 0.01 eV/atom) are applied to the ends of the upper tube to simulate the effect of substrate adhesion.

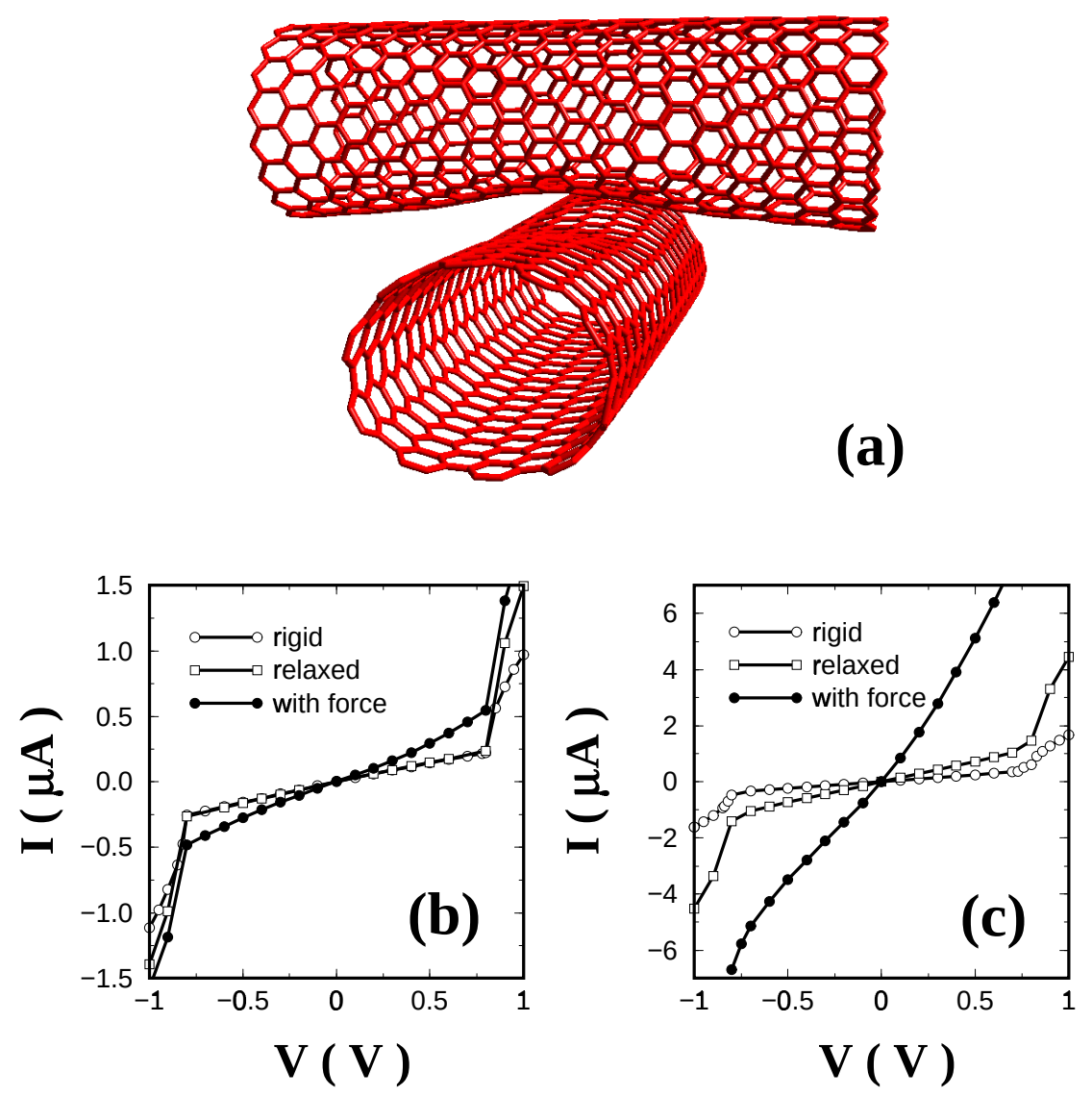

FIG. 4. (a) Four-terminal cross-junction which has two nanotubes perpendicular to each other. The cross-junction is relaxed on a rigid graphite surface. (b) I-V characteristics of a (10,10)-(10,10) (out-of-registry) cross-junction when the tubes are rigid, relaxed without force and relaxed with force applied to the ends of the upper tube. The resistance values are 3.36, 3.21 and 1.66 MΩ for rigid, relaxed without force and relaxed with force cases. (c)I-V characteristics of a (18,0)-(10,10) (in-registry) cross-junction. The resistance value for rigid case is 2.05 MΩ, but decreases to 682 kΩ with relaxation and to 121 kΩ with force.

The conductance and current-voltage characteristics of two junctions are presented in figure 4(b) and (c). In the case of two perpendicular (10,10) tubes, in which the the contact region is incommensurate, the resistance between tubes is 3.36MΩ for rigid nanotubes and decreases to 3.21MΩ when the junction is relaxed. Applying forces further reduces the resistance to 1.66MΩ (Fig.5(a)). In contrast, when tubes are in-registry and the contact region is commensurate, the resistance drops dramatically with relaxation and applying forces. For example, in the case of (18,0)-(10,10) junction the resistance is 2.05 MΩ for rigid tubes but reduced to 682 KΩ after relaxation. When forces are applied the resistance drops to 121 KΩ. The contact resistance we calculated is in agreement with recent experiment [10] which found 90-360 KΩ resistances for metal-metal cross-junctions on a surface. Our results suggest that modest pressure/force can dramatically enhance the inter nanotube transport if the tubes are in-registry.

We have also investigated the dependence of contact resistance on nanotube size and found that the resistance increase with tube diameter. This effect is due to the fact that the two conducting channels of nanotube are coherent states around the whole circumference. Increasing the tube size, though increase the geometrical contact area, in fact reduce the relative weight of conducting channel wave function around the contact. However, we found the effect of relaxation and forces are more dramatic for larger tubes as they are more susceptible to deformation.

In conclusion, we investigated quantum transport properties of intermolecular nanotube contacts. We find that atomic structure in the contact region plays important roles and contact resistance varies strongly with geometry and nanotube chirality. Negative differential resistance is found in nanotube end-end contacts. For cross-junctions, low resistance is achieved when two tubes are in-registry and the contact region is commensurate. Contact resistance can be dramatically decreased by exerting small force/pressure between the tubes if the tubes are in-registry. Significant variation and nonlinearity of contact resistance may lead to new device applications.

## ACKNOWLEDGMENTS

The authors are greatful to S. Paulson, M. R. Falvo, R. Superfine and S. Washburn for many stimulating discussions. This work is supported by U. S. Office of Naval Research (N00014-98-1-0593).